\documentclass[english,aip,apl,preprint,numerical]{revtex4-1}
\usepackage{lmodern}
\usepackage{lmodern}
\usepackage[T1]{fontenc}
\usepackage[latin9]{inputenc}
\usepackage{graphicx}

\makeatletter
\usepackage{babel}

\makeatother

\usepackage{babel}
\begin{document}

\title{Electrical detection of ferromagnetic resonance in ferromagnet/$n$-GaAs
heterostructures by tunneling anisotropic magnetoresistance}

\author{C. Liu}

\affiliation{School of Physics and Astronomy, University of Minnesota, Minneapolis,
Minnesota 55455,~USA}

\author{Y. Boyko}

\altaffiliation{Current address: Department of Physics, University of Maryland, College Park, MD 20742, USA}

\affiliation{School of Physics and Astronomy, University of Minnesota, Minneapolis,
Minnesota 55455,~USA}

\author{C. C. Geppert}

\affiliation{School of Physics and Astronomy, University of Minnesota, Minneapolis,
Minnesota 55455,~USA}

\author{K. D. Christie}

\affiliation{School of Physics and Astronomy, University of Minnesota, Minneapolis,
Minnesota 55455,~USA}

\author{G. Stecklein}

\affiliation{School of Physics and Astronomy, University of Minnesota, Minneapolis,
Minnesota 55455,~USA}

\author{S. J. Patel}

\affiliation{Department of Materials, University of California, Santa Barbara,
California 93106,~USA}

\author{C.~J.~Palmstrøm}

\affiliation{Department of Materials, University of California, Santa Barbara,
California 93106,~USA}

\affiliation{Department of Electrical and Computer Engineering, University of
California, Santa Barbara, California 93106,~USA}

\author{P.~A.~Crowell}

\email{crowell@physics.umn.edu}

\affiliation{School of Physics and Astronomy, University of Minnesota, Minneapolis,
Minnesota 55455,~USA}
\begin{abstract}
We observe a dc voltage peak at ferromagnetic resonance (FMR) in samples
consisting of a single ferromagnetic (FM) layer grown epitaxially
on the $\mathit{n-}$GaAs (001) surface. The FMR peak is detected
as an interfacial voltage with a symmetric line shape and is present
in samples based on various FM/$n$-GaAs hetrostructures, including
Co$_{2}$MnSi/$n$-GaAs, Co$_{2}$FeSi/$n$-GaAs and Fe/$n$-GaAs.
We show that the interface bias voltage dependence of the FMR signal
is identical to that of the tunneling anisotropic magnetoresistance
(TAMR) over most of the bias range. Furthermore, we show how the precessing
magnetization yields a dc FMR signal through the TAMR effect and how
the TAMR phenomenon can be used to predict the angular dependence
of the FMR signal. This TAMR-induced FMR peak can be observed under
conditions where no spin accumulation is present and no spin-polarized
current flows in the semiconductor. 
\end{abstract}
\maketitle
One of the goals of spintronics research is to develop tools for manipulating
electron spins in semiconductors. \cite{Zutic2004} Although many
approaches are based on spin-polarized charge currents, a separate
class of effects is based on the phenomenon of spin pumping, in which
a non-equilibrium spin population is generated by ferromagnetic resonance
(FMR). \cite{Tserkovnyak2002} In the case of metals and semiconductors,
a common method of detecting this effect is to measure the dc voltage
generated by the pumped spin current through the inverse spin Hall
effect. \cite{Saitoh2006,Ando2011,Kurebayashi2011} To correctly interpret
these measurements, it is essential to understand all of the mechanisms
by which the FMR can contribute to the generation of dc voltages.
Among these are anisotropic magnetoresistance (AMR) \cite{Mecking2007}
and the planar Hall effect.\cite{Chen2013,Chen2014}

In this Letter, we report on electrically detected FMR in epitaxial
ferromagnet (FM)/$n$-GaAs (001) heterostructures. The FM/GaAs interfaces
in each of these devices are Schottky tunnel barriers. We find that
the dominant contribution to the electrically detected FMR signal
under reverse and small forward bias current is tunneling anisotropic
magnetoresistance (TAMR). The measured TAMR signal is used to predict
the bias dependence of the FMR signal as well as its dependence on
the magnetic field orientation. The agreement with the predictions
of our model, in which spin transport in the semiconductor plays no
role, is excellent.

The FM/$n$-GaAs heterostructures investigated in this experiment
were grown by molecular beam epitaxy on GaAs (001) substrates. The
growth started with a 500 nm undoped GaAs buffer layer, followed by
2500 nm of Si-doped $n$-GaAs ($n$ = $3-5$$\times$$10^{16}$$\mathrm{cm^{-3}}$).
The junction region consists of a 15 nm $n\rightarrow n^{+}$-GaAs
transition layer followed by 15-18 nm $n^{+}$($5$$\times$$10^{18}$$\mathrm{cm^{-3}}$)
GaAs.\cite{Lou2007} The 5~nm thick FM film is then deposited epitaxially,
followed by 10~nm thick Al and Au capping layers. The FM films studied
are Co$_{2}$MnSi, Co$_{2}$FeSi, and Fe, with deposition temperatures
of $220^{\circ}\mathrm{C}$, $270^{\circ}\mathrm{C}$, and room temperature,
respectively. The first two materials are Heusler alloys that are
promising candidates for spintronics research. \cite{Galanakis2002,Wang2005,Gercsi2006,Farshchi2013}
Devices fabricated from these heterostructures all show non-local
spin valve and Hanle signals in traditional electrical spin injection/detection
measurements at low temperatures. \cite{Lou2007} The FMR signals
discussed in this paper are not strongly temperature dependent, so
only room temperature measurements will be presented.

\begin{figure}
\includegraphics[width=6.5cm]{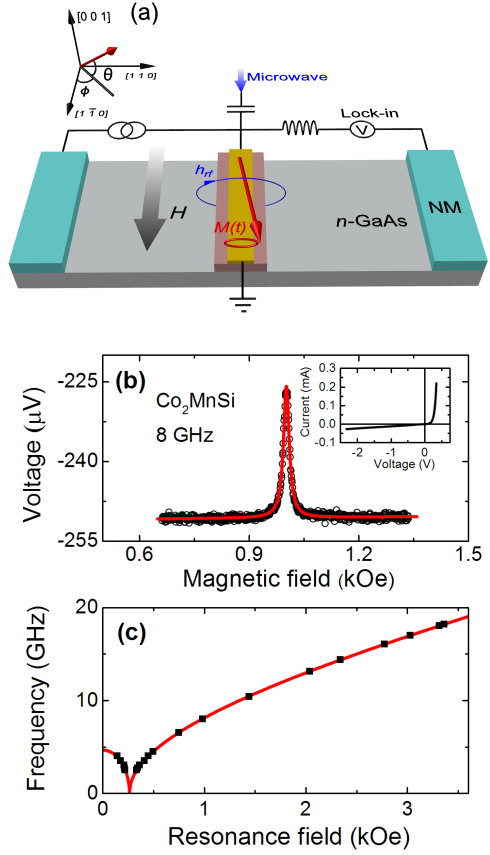} \protect\protect\protect\protect\protect\caption{(color online) (a) Schematic of the measurement geometry. The middle
contact, which has a lateral size of $5\times50$ $\mu m$, is ferromagnetic.
On top of the ferromagnet is a 100 nm thick gold wire. The other two
contacts are fabricated from CuGe. $\mathit{H}$ and $M(t)$ represent
the applied magnetic field and the time-dependent magnetization respectively.
$h_{rf}$ represents the in-plane Oersted field generated by the microwave
current flowing in the gold wire. (b) The dc voltage peak measured
on a Co$_{2}$MnSi sample at 8 GHz under a bias voltage of -0.5V (reverse
bias) as the field is swept through the resonance. The solid line
is a Lorentzian fit. The inset of (b) shows the current-voltage characteristic
of the Co$_{2}$MnSi/$n$-GaAs interface. (c) Experimentally determined
FMR frequency as a function of the magnetic field (solid squares)
applied along the $[1\bar{1}0]$ direction, which is the in-plane
magnetic hard axis. The FMR frequency calculated from the Kittel formula
is shown using the solid line.}

\label{fig:fig1} 
\end{figure}

Figure~\ref{fig:fig1}(a) depicts the measurement geometry for our
experiment, where $\phi$ is the in-plane angle relative to the crystal
axis $[1\bar{1}0]$, which is the in-plane magnetic hard axis, and
$\theta$ is the out-of-plane angle, measured relative to the (001)
plane. For FMR measurements, the magnetic field is applied in the
(001) plane. A dc bias current is combined with the microwave excitation
signal using a bias-T and coupled into the sample using a coaxial
cable. The microwave current passes through a 100 nm thick gold layer
deposited on top of the FM contact, generating an in-plane Oersted
field $h_{rf}$ along the $[110]$ direction. We use a lock-in amplifier
to measure the voltage of FM contact with respect to a CuGe (non-magnetic)
counter electrode.\cite{Aboelfotoh1994} As the magnetic field is
swept through the resonance, a dc voltage peak is measured. Figure~\ref{fig:fig1}(b)
shows the resonance peak for a Co$_{2}$MnSi sample. Similar peaks,
all with a symmetric lineshape, are observed in the other two heterostructures.
By varying the excitation frequency, the FMR frequency can be measured
as a function of magnetic field, as shown using solid squares in Fig.~\ref{fig:fig1}(c).
The frequency calculated from the Kittel formula\cite{Kittel1948}
is shown by the solid curve in Fig.~\ref{fig:fig1}(c). In applying
the Kittel formula, the saturation magnetization $M_{s}$ and uniaxial
anisotropy $K_{u}$ were determined from measurements of the saturation
field along $[001]$ and $[1\bar{1}0]$ directions. 

Because the FMR measurement uses the 3-terminal configuration,\cite{Lou2006}
the observed FMR peak corresponds to a change in the voltage across
the FM/$n$-GaAs interface. Careful characterization of the FM/$n$-GaAs
interface allows us to identify the mechanism responsible for this
FMR peak. In these epitaxally grown samples, a Schottky tunnel barrier
exists at the FM/$n$-GaAs interface.\cite{Schmidt2005} Spin-orbit
interactions due to the Rashba field at the interface as well as the
Dresselhaus field in the tunnel barrier lead to a dependence of the
tunneling resistance on the orientation of the magnetization with
respect to the crystal axes.\cite{Moser2007,Matos-Abiague2009} This
phenomenon is called tunneling anisotropic magnetoresistance (TAMR)
and is present in all of our samples.

\begin{figure}
\includegraphics{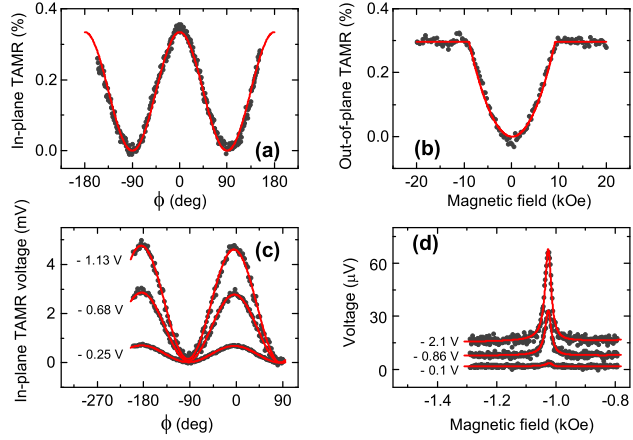} \protect\protect\protect\protect\protect\caption{(color online) (a), (b) The in-plane and out-of-plane TAMR measured
on a Co$_{2}$MnSi sample at bias voltages of -0.5 V and -2.1 V respectively.
In (a) a magnetic field much larger than the in-plane hard-axis saturation
field was used to rotate the magnetization in the film plane. In (b)
a magnetic field normal to the film was applied to gradually align
the magnetization into the out-of-plane direction. (c) The interface
voltage as a function of the in-plane magnetization direction for
different reverse bias voltages. (d) The field-swept FMR peak measured
at different reverse bias voltages.}

\label{fig:fig2} 
\end{figure}

The TAMR effect is shown for a Co$_{2}$MnSi device in Figs.~\ref{fig:fig2}(a)
and (b). The in-plane TAMR $[R(\phi)-R(0)]/R(0)$, where $R(\phi)$
is the interfacial resistance when the magnetization vector is oriented
along $\phi$, is shown as a function of $\phi$ in Fig.~\ref{fig:fig2}(a),
and the out-of-plane TAMR is shown as a function of the out-of-plane
field in Fig.~\ref{fig:fig2}(b). A linear background due to a slight
misalignment of the sample has been subtracted. The observed TAMR
effect in our heterostructures is similar in magnitude to the results
from other studies of FM-GaAs interfaces.\cite{Uemura2011,Moser2007}
The solid line in Fig.~\ref{fig:fig2}(a) is fit using a $\sin^{2}\phi$
function, from which we obtain the magnitude $\Delta R_{i}$ of the
in-plane TAMR. Given the ordinary shape anisotropy of a thin film,
the out-of-plane TAMR should depend quadratically on magnetic field
below saturation. A fit is shown using the solid curve in Fig.~\ref{fig:fig2}(b).
The out-of-plane TAMR $\Delta R_{o}$ is the difference between the
resistances measured at zero field, for which the magnetization lies
along {[}110{]}, and at saturation, for which it lies along {[}001{]}.
The full angular dependence of the TAMR can be written as 
\begin{equation}
R(\phi,\theta)=R(0,0)-\Delta R_{i}\cos^{2}\theta\sin^{2}\phi+\Delta R_{o}^{h}\sin^{2}\theta,\label{eq:TAMR}
\end{equation}
where $\Delta R_{o}^{h}=R(0,\pi/2)-R(0,0)=\Delta R_{\mathrm{\mathit{o}}}-\Delta R_{i}$
is the out-of-plane TAMR measured relative to the $[1\bar{1}0]$ direction.
Similar angular dependencies of the TAMR are observed for any bias
voltage, as shown in Fig.~\ref{fig:fig2}(c) for the in-plane case.
There is a marked similarity in the bias-dependence of the magnitude
of the FMR peak, which is shown in Fig.~\ref{fig:fig2}(d). 
\begin{figure}
\includegraphics[width=7cm]{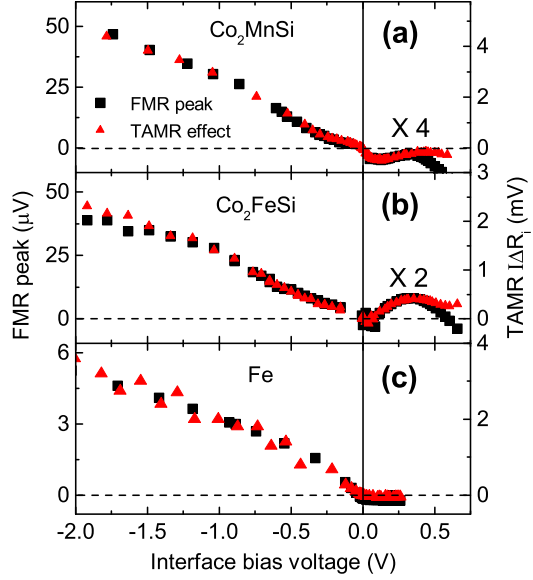} \protect\protect\protect\protect\protect\protect\protect\caption{The magnitudes of the dc voltage peak at FMR (squares, left axis)
and the TAMR voltage $I\Delta R_{i}$ (triangles, right axis) for
the three different FM-GaAs heterostructures at room temperature.
Positive voltages correspond to forward bias (flow of electrons into
the metal) across the Schottky barrier. In (a) and (b), the plots
on the forward bias side are magnified to see the sign difference
of the FMR peak and TAMR effect between the two samples.}

\label{fig:fig3} 
\end{figure}

We now show that the FMR signals in Fig.~\ref{fig:fig1} and Fig.~\ref{fig:fig2}(d)
are due to TAMR. The primary evidence comes from a comparison of the
bias voltage dependence of the magnitude of the FMR peak and the in-plane
TAMR voltage $\Delta V_{i}$. These are shown in Fig.~\ref{fig:fig3}
for all three FM materials. In each case, the FMR and TAMR signals
under reverse bias ($V<0$) are directly proportional to each other.
For clarity, the $y$-scale for forward bias ($V>0$) is magnified
by the factors shown for each sample. A similar scaling between the
TAMR and FMR is observed for small forward bias voltages, although
the proportionality breaks down as the forward bias voltage increases.
This breakdown of scaling between FMR and TAMR is due to the existence
of a spin accumulation, the consequences of which will be discussed
in a future publication.

\begin{figure}
\includegraphics[width=6.5cm]{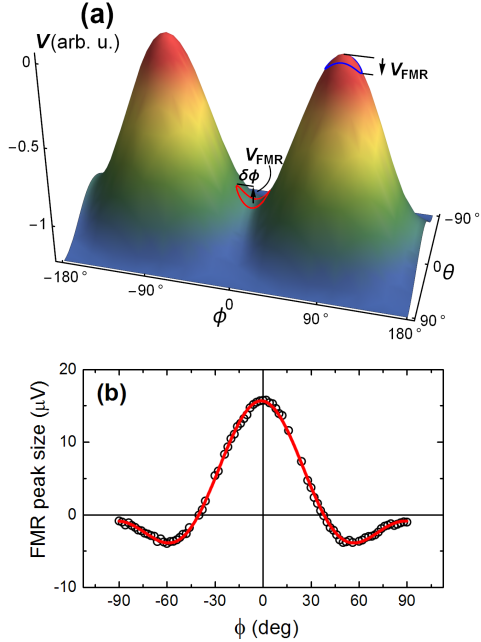} \protect\protect\protect\protect\protect\protect\protect\caption{(color online) (a) The interface voltage as a function of the orientation
($\phi,\theta$) of the magnetization due to the TAMR effect. On resonance
the magnetization traces a trajectory on the 3D surface. (b) The measured
FMR peak size (open circles) as a function of the in-plane angle of
the applied field. The solid line is a fit to Eq.~\ref{eq:VFMR}.}

\label{fig:fig4} 
\end{figure}

In a similar experiment carried out in a waveguide, in which the sample
orientation can be changed with respect to the microwave field, we
find that the peak shape is insensitive to the direction of the microwave
electric field, in contrast to the case of rectification of ordinary
AMR.\cite{Harder2011} This observation implies that the dc voltage
generated by the FMR is sensitive only to the precessing magnetization
and is independent of the microwave current flowing in the FM. The
relevant mechanism is illustrated in Fig.~\ref{fig:fig4}(a), which
shows the TAMR voltage as a function of $\phi$ and $\theta$. When
the FM contact is driven on resonance, the magnetization follows an
elliptical trajectory in $(\phi,\theta)$ space. The resonant trajectory
can be calculated from the known anisotropy surface, and the example
for the case ($\phi=0,\theta=0$) is shown as the solid red curve
in Fig.~\ref{fig:fig4}(a). On average, the TAMR voltage in the presence
of a precessing magnetization \textit{increases} relative to its equilibrium
value at ($\phi=0,\theta=0$). This effect is proportional to the
local curvature of the TAMR surface and the square of the angular
amplitude of precession.

To explore this effect more quantitatively, we investigate the dependence
of the FMR peak on the in-plane orientation of the magnetization at
a fixed reverse bias. This measurement is carried out in a waveguide,
and the orientation of the microwave magnetic field is the same as
in Fig.~\ref{fig:fig1}(a). Figure \ref{fig:fig4}(b) shows the FMR
peak magnitude observed from a Co$_{2}$MnSi sample as a function
of $\phi$. The FMR peak is largest at $\phi=0^{\circ}$ and undergoes
a sign change before approaching zero as $\phi\rightarrow90^{\circ}$.

From the above discussion, we can derive an expression for the magnitude
of the FMR voltage peak as a function of the in-plane angle of the
magnetization. We expand the interface voltage to second order in
small deviations $\delta\phi$ and $\delta\theta$ about their equilibrium
values. We retain only those terms that will not vanish after taking
a time average: 
\begin{eqnarray}
V= & IR(\phi,\theta)=IR(\phi,0)+\left.\frac{1}{2}I\frac{\partial^{2}R(\phi,\theta)}{\partial\phi^{2}}\right|_{\theta=0}(\delta\phi\cos\omega t)^{2}\nonumber \\
 & +\left.\frac{1}{2}I\frac{\partial^{2}R(\phi,\theta)}{\partial\theta^{2}}\right|_{\theta=0}(\delta\theta\sin\omega t)^{2},\label{eq:expansion}
\end{eqnarray}
where $I$ is the interface bias current, $\delta\phi$ and $\delta\theta$
are the in-plane and out-of-plane precession cone angles respectively,
and $\omega$ is the resonance frequency. With the substitution of
the measured $R(\phi,\theta)$ from Eq.~\ref{eq:TAMR} into Eq.~\ref{eq:expansion}
and taking of the time average, we obtain: 
\begin{eqnarray}
\langle V\rangle= & IR(\phi,0)-\frac{1}{2}I\Delta R_{i}\cos2\phi(\delta\phi)^{2}\nonumber \\
+ & \frac{1}{2}(I\Delta R_{i}\sin^{2}\phi+I\Delta R_{o}^{h})(\delta\theta)^{2}.\label{eq:average}
\end{eqnarray}
In Eq.~\ref{eq:average}, the sum of the last two terms, which depend
on the precessional cone angles, is the voltage of the FMR peak. The
precessional cone angles are $\delta\phi=\delta\phi_{0}\cos\phi$
and $\delta\theta=\delta\theta_{0}\cos\phi$, where $\delta\phi_{0}$
and $\delta\theta_{0}$ are the in-plane and out-of-plane angular
amplitudes at $\phi=0^{\circ}$. The factor $\cos\phi$ accounts for
the change in the component of the microwave magnetic field perpendicular
to the magnetization. Finally we obtain: 
\begin{eqnarray}
V_{FMR}(\phi)=-\frac{1}{2}I\Delta R_{i}\cos2\phi\cos^{2}\phi(\delta\phi_{0})^{2}\nonumber \\
+\frac{1}{2}(I\Delta R_{i}\sin^{2}\phi+I\Delta R_{o}^{h})\cos^{2}\phi(\delta\theta_{0})^{2}.\label{eq:VFMR}
\end{eqnarray}

To calculate $V_{FMR}$ from Eq.~\ref{eq:VFMR}, $I\Delta R_{i}$
and $I\Delta R_{o}^{h}$ are obtained from the TAMR measurement. Because
of the shape anisotropy of the thin film, the second term involving
the out-of-plane cone angle in Eq.~\ref{eq:VFMR} is significantly
smaller than the first term. The magnitude of the FMR peak should
therefore be proportional to $I\Delta R_{i}$, as observed in Fig.~\ref{fig:fig3}.
The quadratic dependence on $\delta\phi$ and $\delta\theta$ implies
that the FMR peak should be symmetric, in agreement with experiment.
In Fig.~\ref{fig:fig4}(b) the solid curve is a fit of the in-plane
angle dependence of the FMR signal using Eq.~\ref{eq:VFMR}. The
angular amplitudes $\delta\phi_{0}$ and $\delta\theta_{0}$ are the
only fitting parameters. We find $\delta\phi_{0}=8.4^{\circ}\pm0.3^{\circ}$
and $\delta\theta_{0}=3.7^{\circ}\pm0.2^{\circ}$ for the in-plane
and out-of-plane cone angles respectively. We calculated the dynamical
susceptibility for this sample using the measured saturation magnetization
and anisotropy, from which we find the ratio $\delta\phi_{0}/\delta\theta_{0}\approx2.2$,
in reasonable agreement with the value of 2.3 obtained from the fit
of the angle dependence data in Fig.~\ref{fig:fig4}.

We emphasize that the mechanism discussed in this paper is essentially
a modulation of the tunneling current due to the precession of the
magnetization. This is distinct from spin pumping, in which a spin
current is generated directly by the precessing magnetization. Because
of the significant Schottky tunnel barrier present in these devices,
we expect spin pumping effects to be small. In fact, we have not been
able to observe any inverse spin Hall effect on resonance at zero
bias, in spite of the fact that devices fabricated from the same heterostructures
do function as non-local spin valves. On the other hand, the tunnel
barrier in these samples enhances the TAMR effect. As noted above,
we do observe a significant deviation of the FMR signal from the TAMR
under forward bias voltages, as can be seen in Figs.~\ref{fig:fig3}(a)
and (b). In determining the extent to which these deviations are due
to spin accumulation, a reliable means for separating the TAMR component,
as described here, is essential.

In summary, we have performed electrically detected FMR experiments
on epitaxial FM/$n$-GaAs heterostructures. We observe a strong dc
voltage peak at the FM/$n$-GaAs interface at resonance in a variety
of heterostructures with different ferromagnets. In each case, the
predominant origin of the FMR peak is the tunneling anisotropic magnetoresistance.
This contribution must be considered in any measurement in which the
FM/semiconductor interface is biased.

This work was supported by NSF under DMR-1104951, the MRSEC program
of NSF under DMR 08-19885, and C-SPIN, one of the six centers of STARnet,
a SRC program sponsored by MARCO and DARPA.

 \bibliographystyle{apsrev4-1}
\bibliography{tarmpaper_v9}

\end{document}